\documentclass[conference]{IEEEtran}
\usepackage{blindtext, graphicx}
\usepackage{xcolor}
\usepackage{float}
\usepackage{environ}
\usepackage{datetime}
\usepackage{array}
\usepackage{color}
\usepackage{wasysym}

\usepackage{eurosym}
\usepackage{multirow}
\DeclareGraphicsExtensions{.jpg,.png}
\usepackage[justification=centering]{caption}
\usepackage[ruled,vlined]{algorithm2e}

\begin{document}
%
\title{Efficient Large Scale  Clustering based on Data Partitioning}



 \author{\IEEEauthorblockN{Malika Bendechache}
\IEEEauthorblockA{Insight Centre for Data Analytics
  \\School of Computer Science\\
 University College Dublin\\
 Dublin, Ireland\\
 Email: malika.bendechache@ucdconnect.ie}
 \and
 \IEEEauthorblockN{Nhien-An Le-Khac}
 \IEEEauthorblockA{School of Computer Science\\ 
 University College Dublin\\
 Dublin, Ireland\\
 Email: an.lekhac@ucd.ie}

 \and
 \IEEEauthorblockN{M-Tahar Kechadi}
 \IEEEauthorblockA{Insight Centre for Data Analytics
 \\School of Computer Science\\
 University College Dublin\\
 Dublin, Ireland\\
 Email: tahar.kechadi@ucd.ie}
}

\maketitle
\thispagestyle{plain}
\pagestyle{plain}
\raggedbottom 

\begin{abstract}

  Clustering techniques  are very  attractive for extracting  and identifying  patterns in
  datasets.  However, their  application to very large spatial  datasets presents numerous
  challenges such as high-dimensionality data,  heterogeneity, and high complexity of some
  algorithms. For  instance, some algorithms may  have linear complexity but  they require
  the  domain  knowledge  in  order  to determine  their  input  parameters.   Distributed
  clustering techniques  constitute a  very good  alternative to  the big  data challenges
  (e.g.,Volume, Variety, Veracity, and Velocity).  Usually these techniques consist of two
  phases. The first phase  generates local models or patterns and the  second one tends to
  aggregate the  local results  to obtain  global models.   While the  first phase  can be
  executed in  parallel on each site  and, therefore, efficient, the  aggregation phase is
  complex, time  consuming and  may produce  incorrect and  ambiguous global  clusters and
  therefore incorrect models.

  In this paper we propose a new  distributed clustering approach to deal efficiently with
  both phases; generation of local results and generation of global models by aggregation.
  For the first phase,  our approach is capable of analysing the  datasets located in each
  site using different clustering techniques. The  aggregation phase is designed in such a
  way  that the  final clusters  are compact  and accurate  while the  overall process  is
  efficient in  time and  memory allocation.   For the evaluation,  we use  two well-known
  clustering algorithms; K-Means  and DBSCAN.  One of the key  outputs of this distributed
  clustering technique  is that the number  of global clusters  is dynamic; no need  to be
  fixed in advance.  Experimental results show  that the approach is scalable and produces
  high quality results.

\end{abstract}

\begin{IEEEkeywords}
Big Data, spatial data, clustering, distributed mining, data analysis, k-means, DBSCAN.
\end{IEEEkeywords}

\IEEEpeerreviewmaketitle

\section{Introduction}
\label{sec:Int}
Currently, one of the  most critical data challenges which has a  massive economic need is
how to  efficiently mine and  manage all the  data we have  collected.  This is  even more
critical when the collected data is located on different sites, and are owned by different
organisations ~\cite{Han-06}. This led to the development of distributed data mining (DDM)
techniques  to deal  with huge,  multi-dimensional and  heterogeneous datasets,  which are
distributed over a large number of nodes.  Existing DDM techniques are based on performing
partial analysis on  local data at individual  sites followed by the  generation of global models by aggregating these local results. These two steps are not independent since naive approaches to  local analysis may produce  incorrect and ambiguous global  data models. In order to take advantage  of the mined knowledge at different locations,  DDM should have a
view of the knowledge that not only  facilitates their integration, but also minimises the effect of  the local  results on the  global models. Briefly,  an efficient  management of distributed knowledge is one of the key  factors affecting the outputs of these techniques
~\cite{Aouad-07,L.Aouad-09,Aouad-10b,Whelan-011}. Moreover, the data that is collected and
stored in different  locations using different instruments may have  different formats and
features.  Traditional,  centralised data  mining techniques have  not considered  all the
issues of data-driven applications, such as scalability in both response time and accuracy
of solutions, distribution, and  heterogeneity ~\cite{Bertolotto-07}.  Some DDM approaches
are based  on ensemble learning,  which uses various  techniques to aggregate  the results
~\cite{Lekhac-08},  among the  most cited  in  the literature:  majority voting,  weighted
voting, and stacking ~\cite{Chan-95,Reeves-93}.

DDM is  more appropriate for large scale  distributed platforms, where datasets  are often
geographically distributed and owned by different  organisations. Many DDM methods such as
distributed       association       rules       and       distributed       classification
~\cite{Dietterich-00,Kargupta-00,adamo-12,L.Aouad-09,L.Aouad6-10,Le-Khac-10}  have been
proposed and  developed in  the last  few years.  However, only  a few  researches concern
distributed clustering for analysing large, heterogeneous and distributed datasets. Recent
researches ~\cite{Januzaj-04,Le-Khac-07, L.Aouad3-07} have proposed distributed clustering
approaches based  on the same  2-step process: perform partial  analysis on local  data at
individual sites  and then  aggregate them to  obtain global results.   In this  paper, we
propose a  distributed clustering approach based  on the same 2-step  process, however, it
reduces significantly  the amount of  information exchanged during the  aggregation phase,
and generates automatically the correct number of  clusters.  A case study of an efficient
aggregation phase has been developed on spatial  datasets and proven to be very efficient;
the   data  exchanged   is  reduced   by  more  than $98\%$   of  the   original  datasets
~\cite{Laloux-11}.

The approach can use  any clustering algorithm to perform the  analysis on local datasets.
As can  be seen  in the  following sections, we  tested the  approach with  two well-known
centroid-based and density-based clustering algorithms (K-Means and DBSCAN, respectively),
and  the results  are  of very  high  quality.   More importantly,  this  study shows  how
importance of the local mining algorithms,  as the local clusters accuracy affects heavily
the quality of the final models.

The  rest of  the paper  is organised  as follows:  In the  next section  we will  give an
overview of the  state-of-the-art for distributed data mining and  discuss the limitations
of traditional techniques. Then we will present the proposed distributed framework and its
concepts  in  Section  \ref{sec:DDC}.   In  Section  \ref{sec:DDC-EV},  we  evaluated  the
distributed approach using two well-known algorithms;  K-Means and DBSCAN. We discuss more
experimental  results  showing   the  quality  of  our  algorithm's   results  in  Section
\ref{sec:ExRes}. Finally, we conclude in Section \ref{sec:Con}.

\section{Related Work}
\label{Sec:RW}

DDM techniques can be  divided into two categories based on  the targeted architectures of
computing  platforms~\cite{Zaki-00}. The  first,  based on  parallelism, uses  traditional
dedicated and  parallel machines with  tools for communications between  processors. These
machines are generally called super-computers and are very expensive.  The second category
targets a network  of autonomous machines.  These are called  distributed systems, and are
characterised by  a distributed communication  network connecting low-speed  machines that
can be of different architectures, but  they are very abundant ~\cite{ghosh-14}. The main
goal of the second category of techniques is to distribute the work among the system nodes
and try to minimise  the response time of the whole application.  Some of these techniques
have already been developed and implemented in ~\cite{Aouad-07s,Wu-14}.

However, the  traditional DDM methods  are not always effective,  as they suffer  from the
problem of scaling.  This  has led to the development of techniques  that rely on ensemble
learning ~\cite{Rokach-14,Bauer-99}. These new  techniques are very promising. Integrating
ensemble learning  methods in DDM,  will allow to deal  with the scalability  problem. One
solution to deal with  large scale data is to use parallelism, but  this is very expensive
in terms of communications and processing power. Another solution is to reduce the size of
training sets  (sampling).  Each system node  generates a separate sample.   These samples
will be analysed using a  single global algorithm ~\cite{Zhang-96,Jain-99}.  However, this
technique has  a disadvantage  that the sampling  depends on the  transfer time  which may
impact on the quality of the samples.

Clustering algorithms  can be divided  into two  main categories, namely  partitioning and
hierarchical. Different elaborated taxonomies of  existing clustering algorithms are given
in the literature.  Many parallel clustering  versions based on these algorithms have been
proposed in  the literature  ~\cite{L.Aouad3-07,Dhillon-99, Ester-96,  Garg-06, H.Geng-05,Inderjit-00,   Xu-99}.     These   algorithms   are   further    classified   into   two
sub-categories.  The  first consists  of  methods  requiring  multiple rounds  of  message
passing.  They require  a significant amount of synchronisations.  The second sub-category
consists of methods that build local clustering models  and send them to a central site to
build  global models  ~\cite{Laloux-11}.  In  ~\cite{Dhillon-99} and  ~\cite{Inderjit-00},
message-passing  versions  of  the  widely  used  K-Means  algorithm  were  proposed.   In
~\cite{Ester-96}  and ~\cite{Xu-99},  the authors  dealt with  the parallelisation  of the
DBSCAN density-based clustering algorithm.  In  ~\cite{Garg-06} a parallel message passing
version  of the  BIRCH  algorithm was  presented.  A parallel  version  of a  hierarchical
clustering  algorithm, called  MPC for  Message  Passing Clustering,  which is  especially
dedicated to  Microarray data was introduced  in ~\cite{H.Geng-05}.  Most of  the parallel
approaches need either multiple synchronisation  constraints between processes or a global
view of the dataset, or both ~\cite{L.Aouad3-07}.

Another approach presented  in ~\cite{L.Aouad3-07} also applied a merging  of local models
to create the global models. Current approaches  only focus on either merging local models
or  mining a  set  of local  models  to build  global  ones. If  the  local models  cannot
effectively  represent local  datasets  then  global models  accuracy  will  be very  poor
~\cite{Laloux-11}.   Both  partitioning  and  hierarchical  categories  suffer  from  some
drawbacks.  For  the partitioning class,  K-Means algorithm  needs the number  of clusters
fixed  in  advance,  while in  the  majority  of  cases  $K$ is  not  known.  Furthermore,
hierarchical clustering algorithms  have overcome this limit: they do  not need to provide
the number of clusters as an input parameter, but they must define the stopping conditions
for clustering decomposition, which is not an easy task.

\section{Dynamic Distributed Clustering}
\label{sec:DDC}
We first describe the proposed Dynamic  Distributed Clustering (DDC) model, where the local
models are  based on  the boundaries  of clusters. We  also present  an evaluation  of the
approach with different local clustering techniques including centroid-based (K-Means) and
density-based (DBSCAN) techniques.

\begin{figure}[H]
    \centering
    \begin{center}
    \includegraphics[height=9cm, width=\columnwidth]{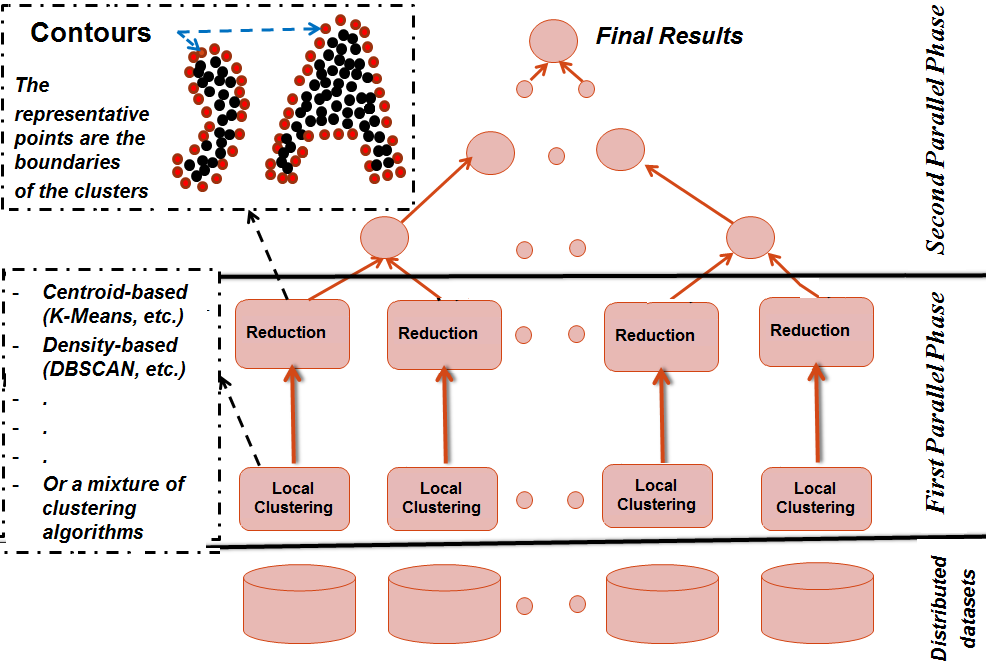}
    \caption{An overview of the DDC Approach.}
    \label{Archi}
     \end{center}
\end{figure}

The DDC  approach includes two main  steps.  In the first  step, as usual, we  cluster the
datasets located on  each node of the  system and select good  local representatives. This
phase is executed in parallel without communications  between the nodes.  In this phase we
can reach a super  speed-up. The next phase, however, collects the  local models from each
node and affects them to some special nodes  in the system called leaders. The leaders are
elected  according to  some  characteristics  such as  their  capacity, processing  power,
connectivity,  etc. The  leaders are  responsible for  merging and  regenerating the  data
objects based on the local cluster representatives. The purpose of this step is to improve
the quality of  the global clusters, as  usually the local clusters do  not contain enough
important information.

\subsection{\textbf{Local Models}}

The  local clusters  are  highly dependent  on  the clustering  techniques  used by  their
corresponding nodes. For instance, for spatial datasets, the shape of a cluster is usually
dictated by  the technique used to  obtain them.  Moreover, this  is not an issue  for the
first phase,  as the  accuracy of  a cluster  affects only  the local  results of  a given
node. However, the second  phase requires sending and receiving all  local clusters to the
leaders. As the  whole data is very  large, this operation will saturate  very quickly the
network. So, we must avoid sending all the original data through the network. The key idea
behind the  DDC approach is to  send only the cluster's  representatives, which constitute
between  $1\%$ and  $2\%$ of  the total  size of  the data.   The cluster  representatives
consist of the internal data representatives plus the boundary points of the cluster.
 
There are  many existing data  reduction techniques in the  literature.  Many of  them are
focusing only on  the dataset size i.e.,  they try to reduce the  storage capacity without
paying attention to the knowledge contained  in the data.  In ~\cite{N-A-10}, an efficient
reduction technique has been proposed; it is based on density-based clustering algorithms.
Each  cluster  is  represented  by  a   set  of  carefully  selected  data-points,  called
representatives.   However, selecting  representatives is  still a  challenge in  terms of
quality and size ~\cite{Januzaj-04,Laloux-11}.
 
The best way  to represent a spatial cluster is  by its shape and density. The  shape of a
cluster   is  represented   by  its   boundary   points  (called   contour)  (see   Figure
\ref{Archi}). Many algorithms for  extracting the boundaries from a cluster  can  be found 
in the  literature  ~\cite{M.J-04,A.Ray-97,M.Melkemi-00, Edelsbrunner-83,A.Moreira-07}. We
used the algorithm  proposed in ~\cite{M.Duckhama-08}, which is based  on triangulation to
generate the cluster boundaries.  It is an efficient algorithm for constructing non-convex
boundaries.  The algorithm is able to accurately characterise the shape of a wide range of
different   point  distributions   and   densities  with   a   reasonable  complexity   of
$\mathcal{O}(n\log{n})$.

\subsection{\textbf{Global Models}}

The global models (patterns) are generated during  the second phase of the DDC. This phase
is  also  executed  in  a  distributed  fashion  but,  unlike  the  first  phase,  it  has
communications overheads.  This phase consists  of two main  steps, which can  be repeated
until  all the  global clusters  were  generated. First,  each leader  collects the  local
clusters of its neighbours.   Second, the leaders will merge the  local clusters using the
overlay technique. The process  of merging clusters will continue until  we reach the root
node. The root node will contain the global clusters (see Figure \ref{Archi}).

Note that, this phase  can be executed by a clustering algorithm, which can be the same as
in the first phase  or completely different one. This approach belongs  to the category of
hierarchical clustering.

As  mentioned above,  during the  second phase,  communicating the  local clusters  to the
leaders  may generate a huge overhead. Therefore, the  objective is  to minimise  the data
communication and computational  time, while getting accurate  global results.
In DDC we  only exchange the boundaries  of the clusters, instead of  exchanging the whole
clusters between the system nodes.

In the following, we  summarise the steps of the DDC approach for  spatial data. The nodes
of the distributed computing system are organised following a tree topology.
\begin{enumerate}
   \item Each node is  allocated a dataset representing a portion of the  scene or of the
         overall dataset.  
   \item Each leaf node executes a local clustering algorithm with its own input 
         parameters.
   \item Each node  shares its  clusters with  its neighbours  in order  to form  larger
         clusters using the overlay technique. 
   \item The leader nodes contain the results of their groups. 
   \item Repeat 3 and 4 until all global clusters were generated.
\end{enumerate}

Note that the DDC  approach suits very well the MapReduce framework,  which is widely used
in cloud computing systems. 

\section{DDC Evaluation and Validation}
\label{sec:DDC-EV}
In  order  to evaluate  the  performance  of the  DDC  approach ,  we use  different  local
clustering algorithms.   In this paper we  use a centroid-based algorithm  (K-Means) and a
density-based Algorithm (DBSCAN).
  
\subsection{DDC-K-Means }
\label{DDC-K}
Following the  general structure  of the  approach described above,  the DDC  with K-Means
(DDC-K-Means) is characterised  by the fact that  in the first phase,  called the parallel
phase, each node $N_i$  of the system executes the K-Means algorithm  on its local dataset
to produce $L_i$ local  clusters and calculate their contours. The rest  of the process is
the same as described above.  In other  words, the second phase consists of exchanging the
contours located in each node with its neighbourhood nodes.  Each leader attempts to merge
overlapping contours  of its group.   Therefore, each  leader generates new  contours (new
clusters). The merge procedure will continue until there is no overlapping contours.

It has been  shown in \cite{M-Bendechache-15} that DDC-K-Means  dynamically determines the
number of the clusters without a priori  knowledge about the data or an estimation process
of the number of the clusters.  DDC-K-Means was also compared to two well-known clustering
algorithms:  BIRCH and  CURE. The  results  showed that  the quality  of the  DDC-K-Means'
clusters is much better that the ones generated by both BIRCH and CURE.  As expected, this
approach runs much faster than the two other algorithms; BIRCH and CURE.

To summarise, DDC-K-Means  does not need the number  of global clusters to be  given as an
input. It  is calculated dynamically. Moreover,  each local clustering $L_i$  with K-Means
needs  $K_i$ as  a  parameter, which  is  not necessarily  the exact  $K$  for that  local
clustering. Let $\tilde{K_i}$ be the exact number of local clusters in the node $N_i$, all
it is required  is to set $K_i$ such  that $K_i > \tilde{K_i}$. This is  much simpler than
giving $K_i$,  especially when  we do not  have enough knowledge  about the  local dataset
characteristics. Nevertheless, it  is indeed better to  set $K_i$ as close  as possible to
$\tilde{K_i}$ in order to reduce the processing  time in calculating the contours and also
merging procedure.

\begin{figure*}[!ht]
\centering
\begin{tabular}{c | c | c | c}
    Original Dataset & BIRCH & CURE & DDC-K-Means \\ \hline \\
    \includegraphics[width=0.18\textwidth, height=0.35\textwidth]{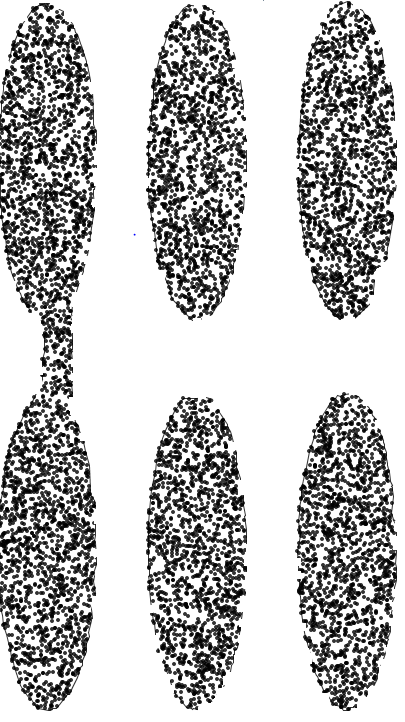} &
    \includegraphics[width=0.18\textwidth, height=0.35\textwidth]{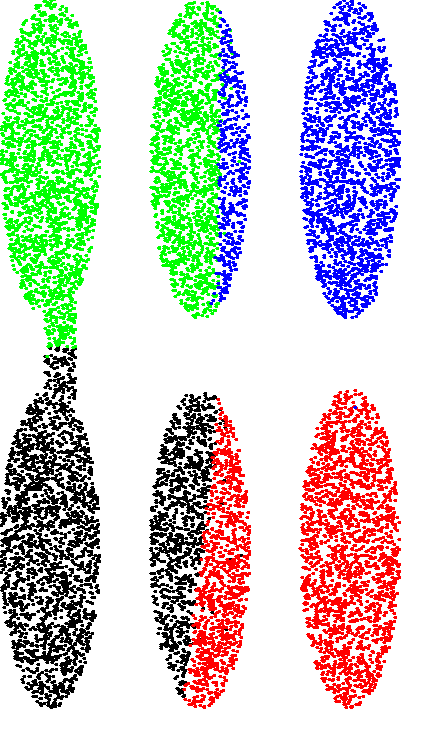} &
    \includegraphics[width=0.18\textwidth, height=0.35\textwidth]{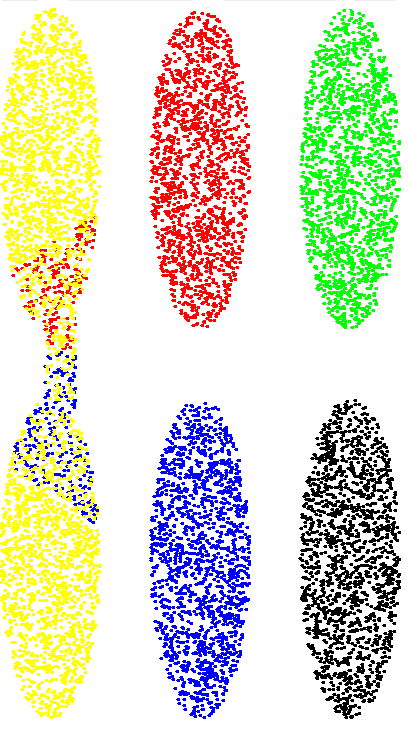} &
    \includegraphics[width=0.18\textwidth, height=0.35\textwidth]{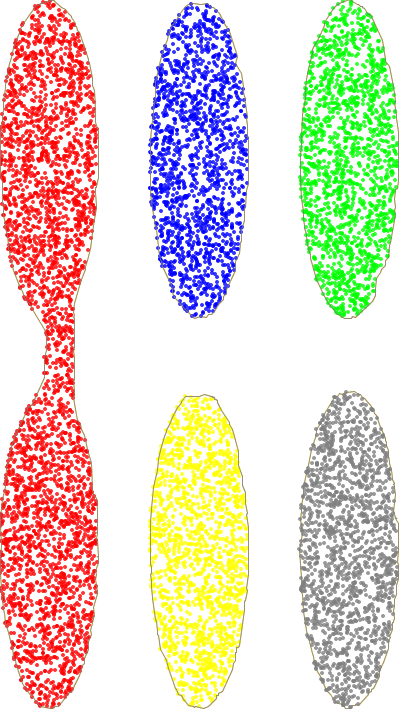} 
 \\ 
 \multicolumn{4}{c}{T1}
 \\ 
    \includegraphics[width=0.18\textwidth, height=0.18\textwidth]{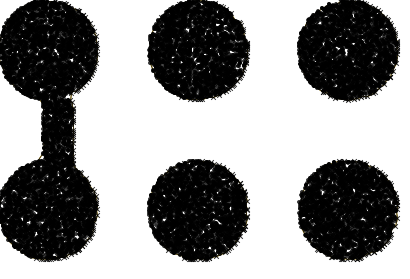} &
    \includegraphics[width=0.18\textwidth, height=0.18\textwidth]{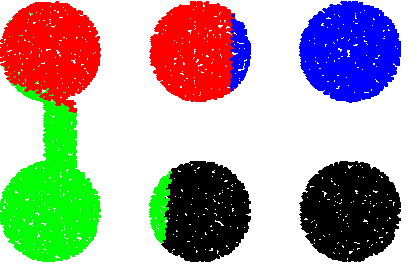} &
    \includegraphics[width=0.18\textwidth, height=0.18\textwidth]{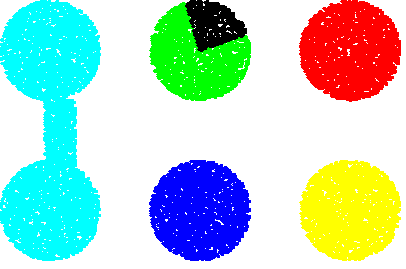} &
    \includegraphics[width=0.18\textwidth, height=0.18\textwidth]{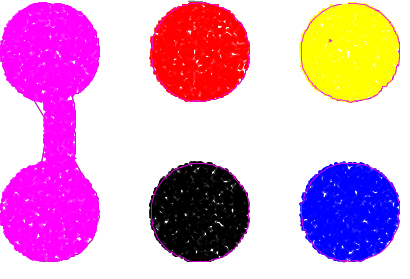} 
 \\ 
  \multicolumn{4}{c}{T2}
 \\ 
    \includegraphics[width=0.18\textwidth, height=0.18\textwidth]{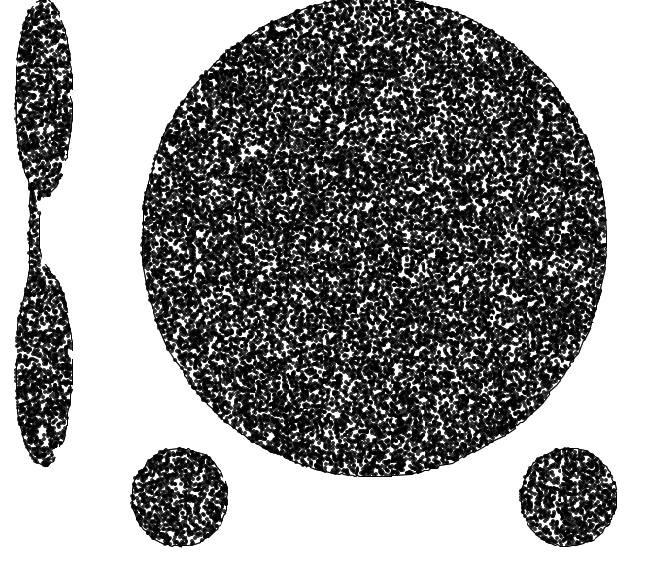} &
    \includegraphics[width=0.18\textwidth, height=0.18\textwidth]{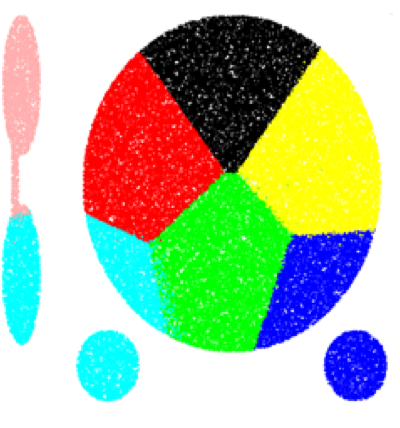} &
    \includegraphics[width=0.18\textwidth, height=0.18\textwidth]{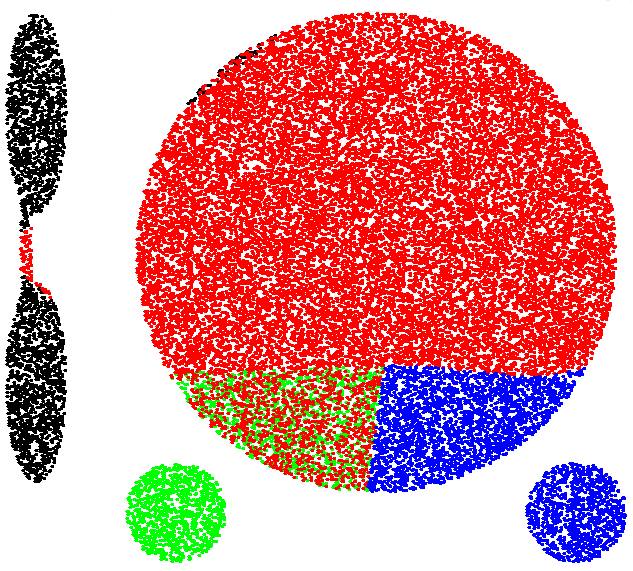} &
    \includegraphics[width=0.18\textwidth, height=0.18\textwidth]{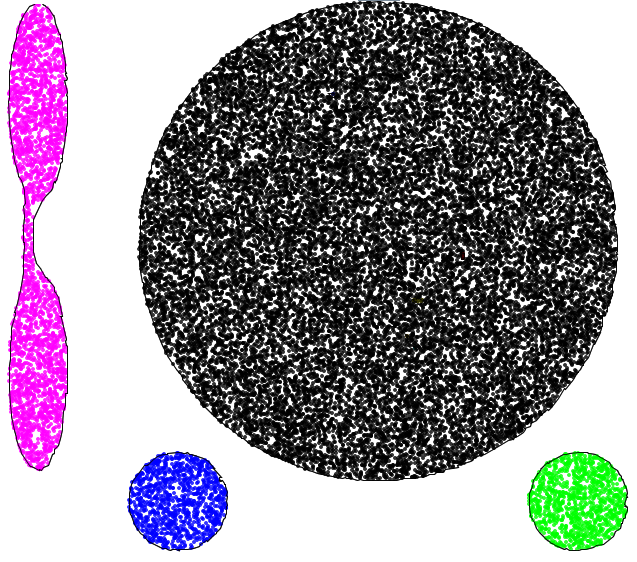} 
 \\ 
  \multicolumn{4}{c}{T3}
 \\ 
    \includegraphics[width=0.18\textwidth, height=0.18\textwidth]{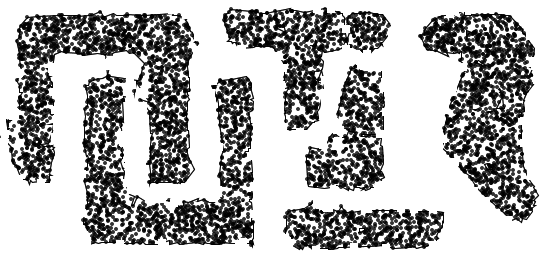} &
    \includegraphics[width=0.18\textwidth, height=0.18\textwidth]{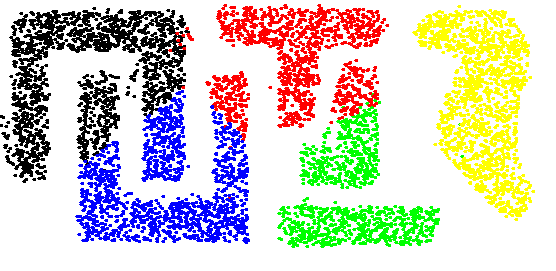} &
    \includegraphics[width=0.18\textwidth, height=0.18\textwidth]{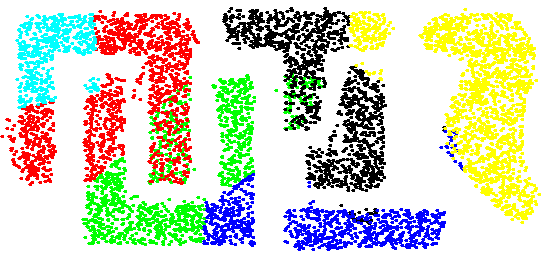} &
    \includegraphics[width=0.18\textwidth, height=0.18\textwidth]{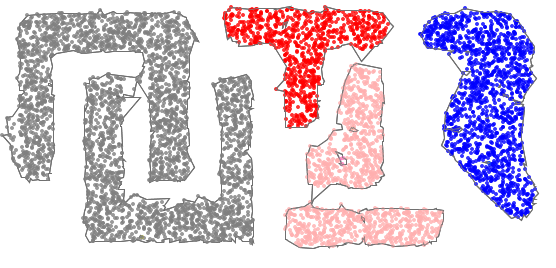} 
 \\ 
  \multicolumn{4}{c}{T4}
 \\ 
    \includegraphics[width=0.18\textwidth, height=0.18\textwidth]{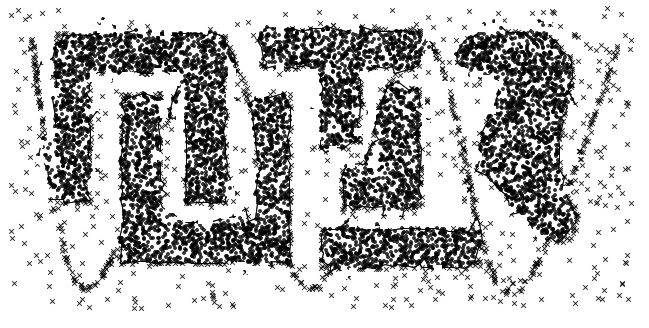} &
    \includegraphics[width=0.18\textwidth, height=0.18\textwidth]{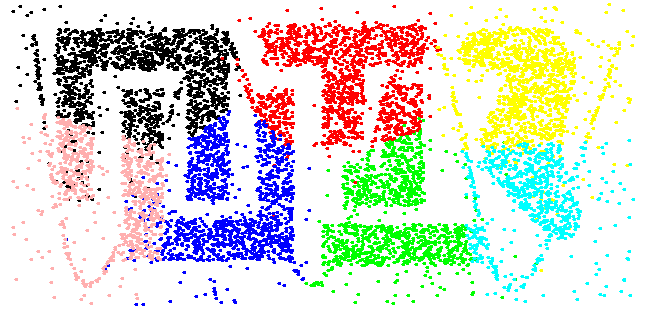} &
    \includegraphics[width=0.18\textwidth, height=0.18\textwidth]{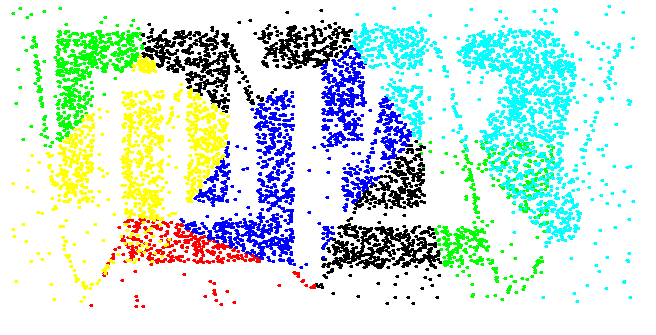} &
    \includegraphics[width=0.18\textwidth, height=0.18\textwidth]{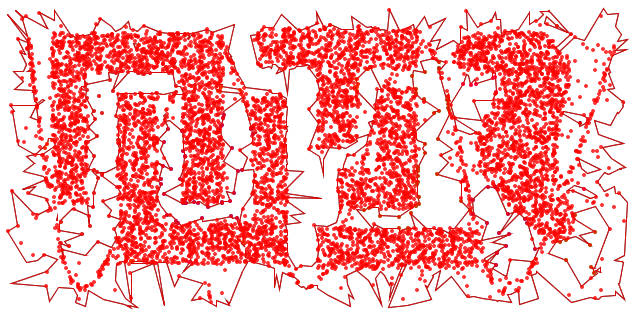} 
\\ 
 \multicolumn{4}{c}{T5 (T4 with noise )}
 \\ 
\end{tabular}
\caption{Comparing the clusters generated across different datasets}
\label{FIG1}
\end{figure*}

Figure  \ref{FIG1}  shows a  comparative  study  between  DDC-K-Means and  two  well-known
algorithms; BIRCH and CURE.  The experiments use five datasets ($T_1, T_2, T_3, T_4, T_5$)
described in Table~\ref{table1}. Note that $T_5$ is the same dataset as $T_4$ for which we
removed the noise.  Each color represents a separate cluster.

As we can see, DDC-K-Means successfully generates the final clusters  for the first three
datasets ($T_1,  T_2$ and  $T_3$), whereas BIRCH  and CURE fail  to generate  the expected
clusters for all the datasets ($T_1, T_2, T_3, T_4$ and $T_5$).
However, DDC-K-Means  fails to  find good clusters  for the two  last datasets  ($T_4$ and
$T_5$), this is due to the fact that the K-Means algorithm tends to work with convex shape
only, because it  is based on the  centroid principle to generate  clusters.  Moreover, we
can also notice that the results of DDC-K-Means are even worse with dataset which contains
noise ($T_5$). In  fact it returns the whole  dataset with the noise as  one final cluster
for each  dataset (see  Figure \ref{FIG1}).  This  is because K-Means  does not  deal with
noise.

\subsection{DDC with DBSCAN}

While, the DDC-K-Means performs much better  than some well-known clustering algorithms on
spatial datasets, it still can not deal with all kinds of datasets; mainly with non-convex
shapes.  In addition,  DDC-K-Means  is  very sensitive  to  noise.  Therefore, instead  of
K-Means, we use another clustering algorithm for spatial datasets, which is DBSCAN. DBSCAN
is summarised below.
 
\subsubsection{\textbf{DBSCAN}}

DBSCAN  (Density-Based spatial  Clustering of  Applications  with Noise)  is a  well-known
density based clustering  algorithm capable of discovering clusters  with arbitrary shapes
and eliminating noisy  data ~\cite{M.Ester-96}.  Briefly, DBSCAN  clustering algorithm has
two main parameters: the radius $Eps$ and minimum points $MinPts$.  For a given point $p$,
the  $Eps$\_{\em neighbourhood} of $p$  is the  set of  all the  points around  $p$ within
distance  $Eps$.  If  the number  of points  in the  $Eps$\_{\em neighbourhood} of $p$  is
smaller than $MinPts$, then  all the points in this set, together with  $p$, belong to the
same cluster. More details can be found in ~\cite{M.Ester-96}.

Compared with other popular clustering methods such as K-Means ~\cite{T.Kanungo-02}, BIRCH
~ \cite{Tian-96}, and STING ~\cite{W.Wang-97}, DBSCAN has several key features.  First, it
groups data into clusters  with arbitrary shapes.  Second, it does  not require the number
of the  clusters to be  given as an  input.  The number of  clusters is determined  by the
nature of the data and the values of  $Eps$ and $MinPts$.  Third, it is insensitive to the
input  order of  points in  the dataset.  All  these features  are very  important to  any
clustering algorithm.
 
 \subsubsection{\textbf{DBSCAN Complexity}}
\label{DBCom}
 DBSCAN visits each point of the dataset,  possibly multiple times (e.g., as candidates to
 different clusters). For practical considerations, however, the time complexity is mostly
 governed by the number of regionQuery invocations. DBSCAN executes exactly one such query
 for each point, and  if an indexing structure is used that executes a neighbourhood query
 in $\mathcal{O}(\log  n)$, an overall  average complexity  of $\mathcal{O}(n \log  n)$ is
 obtained  if the  parameter $Eps$  is chosen  in a  meaningful way,  (i.e., such  that on
 average  only  $  \mathcal{O}(\log n)$  points  are  returned).  Without  the use  of  an
 accelerating index structure, or on degenerated  data (e.g., all points within a distance
 less than  $Eps$), the worst case  run time complexity remains  $\mathcal{O} (n^2)$.  The
 distance matrix  of size $\mathcal{O}((n^2-n/2))$  can be materialised to  avoid distance
 re-computations, but this  needs $\mathcal{O}(n^2)$ of memory, whereas a non-matrix based
 implementation of DBSCAN only needs $O(n)$ of memory space.

\subsubsection{\textbf{DDC-DBSCAN Algorithm}}

The approach remains the same (as explained in Section \ref{sec:DDC}), the only difference
is at local level (See Figure \ref{Archi}). Where, instead of using K-Means for processing
local clusters, we use  DBSCAN. Each node ($n_i$) executes DBSCAN on  its local dataset to
produce $K_i$  local clusters. Once  all the local  clusters are determined,  we calculate
their contours.   These contours will  be used  as representatives of  their corresponding
clusters.

\begin{algorithm}[!htb]
  \SetKwInOut{Input}{input}\SetKwInOut{Output}{output}
  \Input {$X_i$: Dataset Fragment, $Eps_i$: Distance $Eps_i$ for $Node_i$, $MinPts_i$: 
    minimum points contain clusters generated by  $Node_i$, $D$: tree degree, $L_i$: Local
    clusters generated by $Node_i$} 
  \Output {$K_g$: Global Clusters (global results)}
  \BlankLine
  \BlankLine
  $level = tree height$\;
  
  \begin{enumerate}
  \item DBSCAN($X_i$. $Eps_i$, $MinPts_i$)\;  \tcp{$Node_i$ executes DBSCAN locally.}
  \item Contour($L_i$)\; \tcp{$Node_i$ executes the Contour algorithm to generate the boundary 
        of each local cluster.} 
  \item $Node_i$ joins a group $G$ of $D$ elements\; \tcp{$Node_i$ joins its neighbourhood} 
  \item Compare cluster of $Node_i$  to other node's clusters in the same group\;
    \tcp{look for overlapping between clusters.} 
  \item j = ElectLeaderNode()\;
    \tcp{Elect a node which will merge the overlapping clusters.} 
  \end{enumerate}
  \If {$i<>j$}{
    Send (contour i, j)\;
  }
  \Else 
  {
    \If {$level>0$}{
      $level$ - - \;
      Repeat 3, 4, and 5 until level=0\;
    }
    \Else
    {
      \Return ($K_g$: $Node_i$' clusters)\;
    }}
  \caption{DDC with DBSCAN.}
  \label{clustering}
\end{algorithm}

The  second phase  consists of  exchanging  the contours  located  in each  node with  its
neighbourhood  nodes.  This  will allow  us to  identify overlapping  contours (clusters).
Each leader attempts  to merge overlapping contours of its  group.  Therefore, each leader
generates new contours (new clusters). We repeat  the second and third steps till we reach
the root node.   The sub-clusters aggregation is  done following a tree  structure and the
global results are located in the top level  of the tree (root node). The algorithm pseudo
code is given in Algorithm~\ref{clustering}.

\begin{figure}[H]
  \centering
  \begin{center}
 
  \includegraphics[height=11cm, width=\columnwidth]{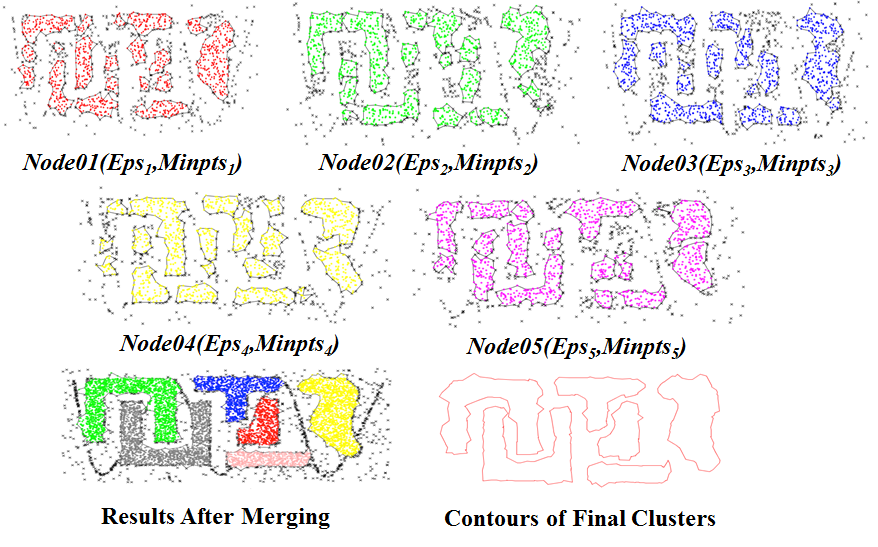}
  \caption{Example of DDC-DBSCAN execution.}
   
  \label{DBS}
  \end{center}
\end{figure}

Figure  \ref{DBS} illustrates  an  example  of DDC-DBSCAN.   Assume  that the  distributed
computing platform contains five Nodes ($N=5$).   Each Node executes DBSCAN algorithm with
its local  parameters ($Eps_i$, $MinPts_i$)  on its  local dataset. As  it can be  seen in
Figure \ref{DBS} the new approach returned exactly  the right number of clusters and their
shapes. The approach is insensitive to the way the original data was distributed among the
nodes. It  is also insensitive to  noise and outliers. As  we can see, although  each node
executed DBSCAN locally with different parameters.  The global final clusters were correct
even on the noisy dataset ($T_5$) (See Figure \ref{DBS}).

\section{Experimental Results}
\label{sec:ExRes}

In this section, we  study the performance of the DDC-DBSCAN  approach and demonstrate its
effectiveness compared to BIRCH, CURE and  DDC-K-Means. We choose these algorithms because
either  they are  in the  same category  as the  proposed technique,  such as  BIRCH which
belongs to hierarchical  clustering category, or have an  efficient optimisation approach,
such as CURE.

\textbf{BIRCH}: We used the BIRCH implementation provided in \cite{Tian-96}. It performs a
pre- clustering  and then  uses a centroid-based  hierarchical clustering  algorithm. Note
that the time and  space complexity of this approach is quadratic to  the number of points
after  pre-clustering.  We  set  its  parameters   to  the  default  values  suggested  in
\cite{Tian-96}.

\textbf{CURE}:  We used  the implementation  of  CURE provided  in \cite{Sudipto-01}.  The
algorithm uses  representative points  with shrinking  towards the  mean. As  described in
\cite{Sudipto-01},  when  two  clusters  are  merged   in  each  step  of  the  algorithm,
representative points  for the new merged  cluster are selected  from the ones of  the two
original clusters rather than all the points in the merged clusters.

\subsection{Experiments}

We run experiments with different datasets.  We  used six types of datasets with different
shapes and sizes.  The  first three datasets ($T_1,T_2,$ and $T_3$)  are the same datasets
used to  evaluate DDC-K-Means. The  last three datasets ($T_4,  T_5,$ and $T_6$)  are very
well-known  benchmarks  to evaluate  density-based  clustering  algorithms.  All  the  six
datasets are summarised in Table \ref{table1}.  The  number of points and clusters in each
dataset is also  given. These six datasets contain  a set of shapes or  patterns which are
not easy to extract with traditional techniques.

\begin{table}[htb]
  \caption{The datasets used to test the algorithms.}
  \centering 
  \begin{center}
    \begin{tabular}{|c|c|c|c|c|}
      \hline
      \textbf{Type}  & \textbf{Dataset} & \textbf{Description} & \textbf{\#Points} & \textbf{\#Clusters} \\ \hline
      \multirow{3}{*}{\textbf{Convex}}  & T1   & \begin{tabular}[c]{@{}c@{}}Big oval\\ (egg shape)\end{tabular}                     & 14,000    & 5  \\ \cline{2-5} & T2   & \begin{tabular}[c]{@{}c@{}}4 small circles\\ and 2 small circles linked\end{tabular}   & 17,080   & 5    \\ \cline{2-5} & T3   & \begin{tabular}[c]{@{}c@{}}2 small circles,\\ 1 big circle\\ and 2 linked ovals\end{tabular} & 30,350     & 4 \\ \hline
      \multirow{3}{*}{\textbf{\begin{tabular}[c]{@{}c@{}}Non-Convex\\ with Noise\end{tabular}}} & T4 & \begin{tabular}[c]{@{}c@{}}Different shapes\\ including noise\end{tabular} & 8,000    & 6     \\ \cline{2-5}  & T5   & \begin{tabular}[c]{@{}c@{}}Different shapes, with \\ some clusters surrounded\\  by others\end{tabular} & 10,000   & 9 \\ \cline{2-5}  & T6 & Letters with noise  & 8,000    & 6    \\ \hline
    \end{tabular}
    \label{table1}
  \end{center}
\end{table}

\subsection{Quality of Clustering}
\label{QC}
\begin{figure*}[!ht]
\centering
\begin{tabular}{c | c | c |c}
    BIRCH &    CURE &    DDC-K-Means&    DDC-DBSCAN \\ \hline \\

    \includegraphics[width=0.22\textwidth, height=0.18\textwidth]{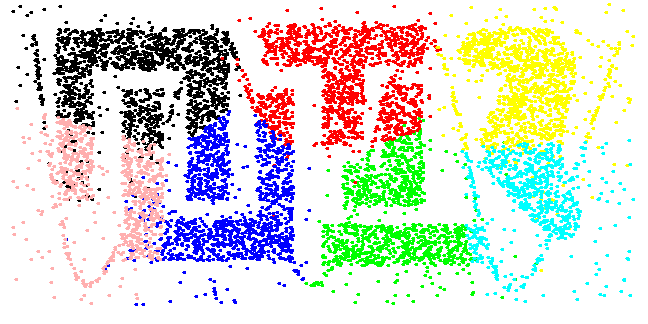} &
    \includegraphics[width=0.22\textwidth, height=0.18\textwidth]{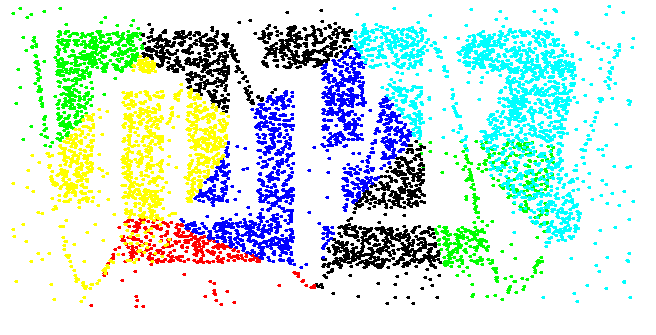} &
    \includegraphics[width=0.22\textwidth, height=0.18\textwidth]{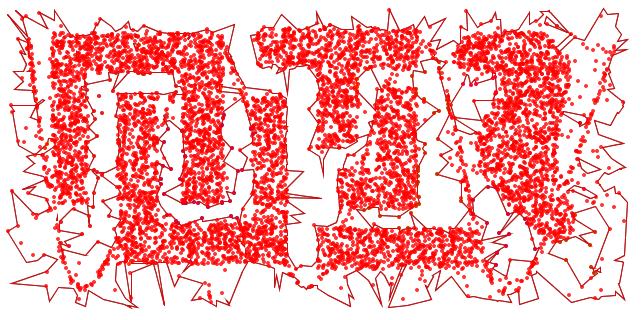} &
     \includegraphics[width=0.22\textwidth, height=0.18\textwidth]{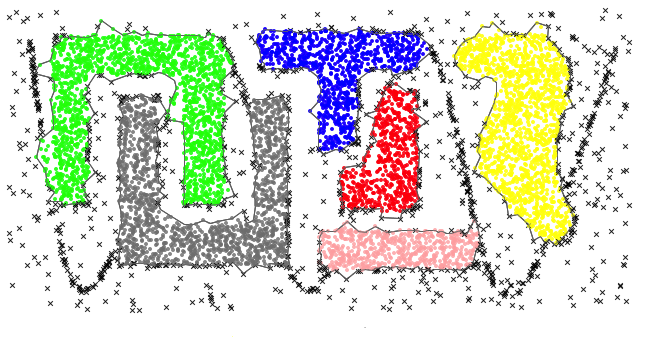}
 \\ 
  \multicolumn{4}{c}{T4}
 \\ 
    \includegraphics[width=0.22\textwidth, height=0.18\textwidth]{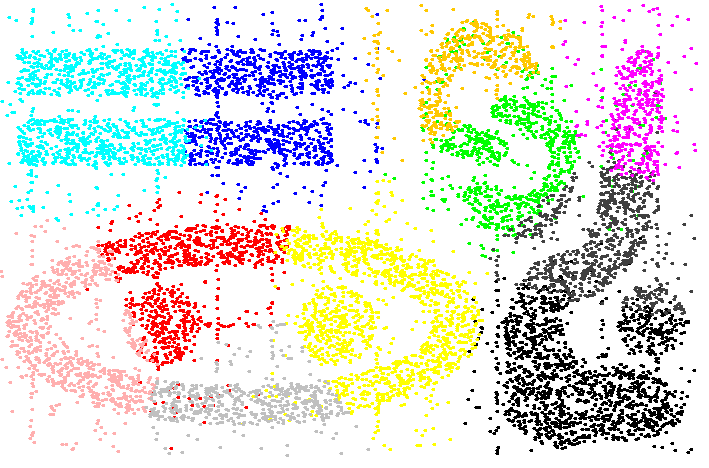} &
    \includegraphics[width=0.22\textwidth, height=0.18\textwidth]{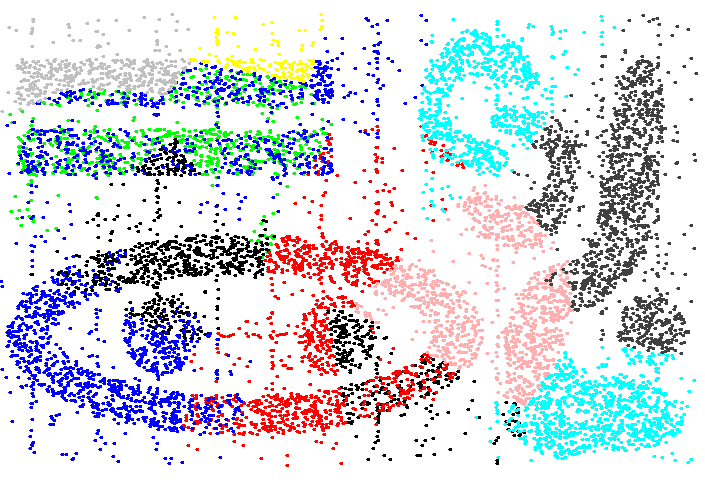} &
    \includegraphics[width=0.22\textwidth, height=0.18\textwidth]{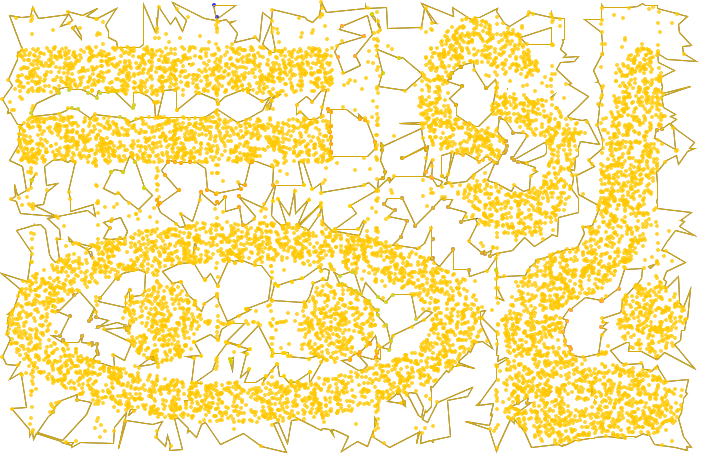} &
     \includegraphics[width=0.22\textwidth, height=0.18\textwidth]{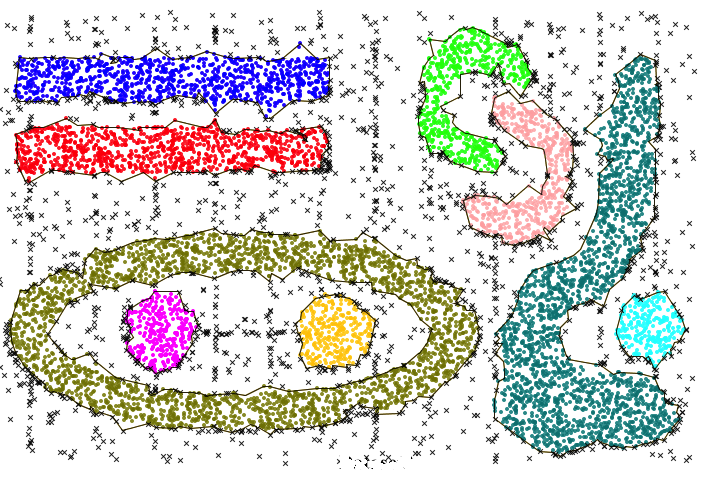}
\\ 
 \multicolumn{4}{c}{T5}
 \\
    \includegraphics[width=0.22\textwidth, height=0.18\textwidth]{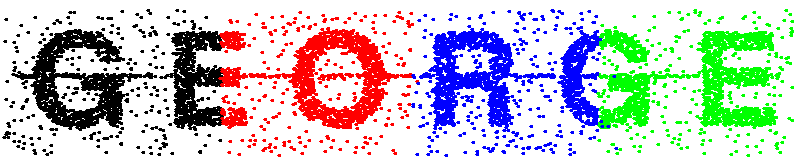} &
    \includegraphics[width=0.22\textwidth, height=0.18\textwidth]{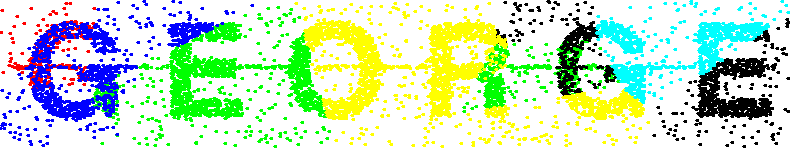} &
    \includegraphics[width=0.22\textwidth, height=0.18\textwidth]{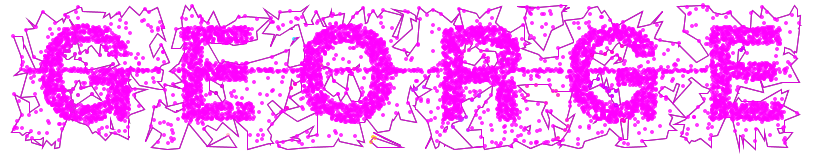} &
     \includegraphics[width=0.22\textwidth, height=0.18\textwidth]{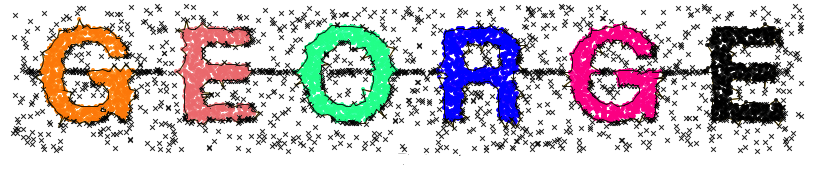}
\\ 
 \multicolumn{4}{c}{T6}
 \\ 
\end{tabular}
\caption{Clusters generated form non-convex shapes and noisy datasets.}
\label{Nonesph}
\end{figure*}

We run the four  algorithms on the six datasets in order to  evaluate the quality of their
final clusters. In the case of the DDC approach we took a system that contains five nodes,
therefore, the  results shown  are the  aggregation of the  five local  clustering. Figure
\ref{SphSh} shows the returned  clusters by each of the four  algorithms for convex shapes
of  the clusters  ($T_1, T_2,$  and  $T_3$) and  Figure \ref{Nonesph}  shows the  clusters
returned for non-convex shapes of the clusters  with noise ($T_4, T_5,$ and $T_6$). We use
different colours to show the clusters returned by each algorithm.


\begin{table*}[!ht]
\centering
\begin{tabular}{c | c | c |c}
    BIRCH &    CURE &    DDC-K-Means&    DDC-DBSCAN \\ \hline \\

    \includegraphics[width=0.22\textwidth, height=0.35\textwidth]{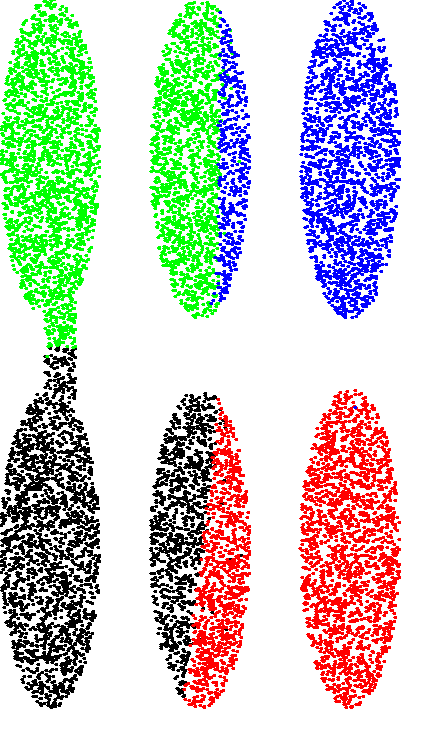} &
    \includegraphics[width=0.22\textwidth, height=0.35\textwidth]{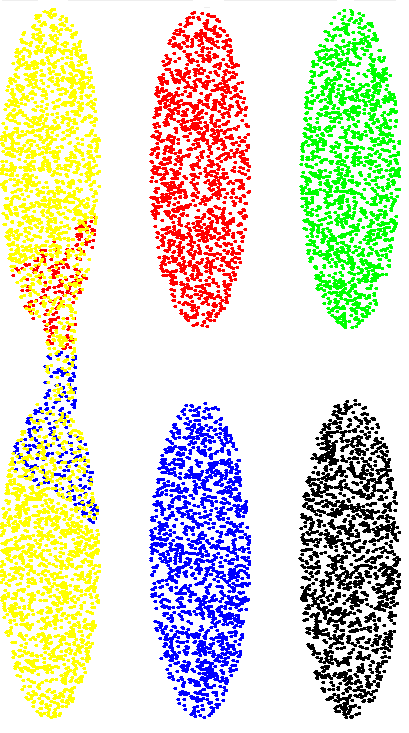} &
    \includegraphics[width=0.22\textwidth, height=0.35\textwidth]{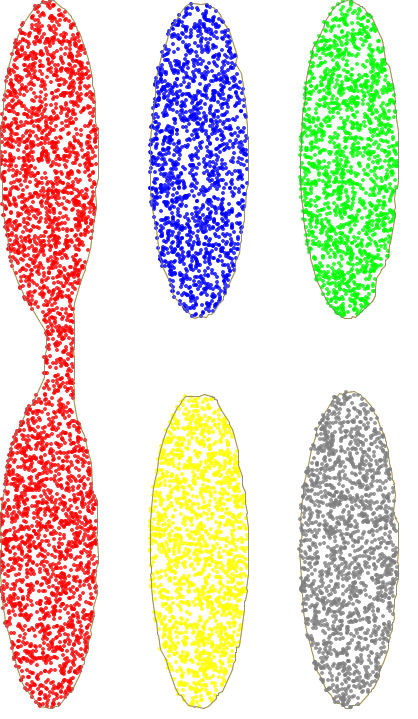} & 
    \includegraphics[width=0.22\textwidth, height=0.35\textwidth]{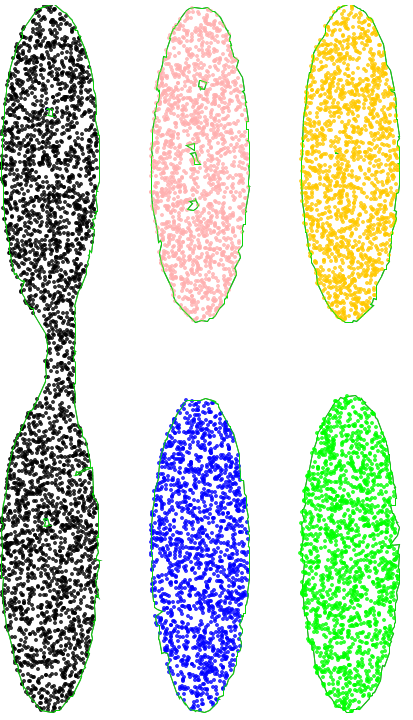} 
 \\ 
 \multicolumn{4}{c}{T1}
 \\ 
    \includegraphics[width=0.22\textwidth, height=0.18\textwidth]{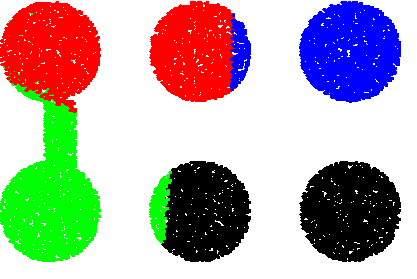} &
    \includegraphics[width=0.22\textwidth, height=0.18\textwidth]{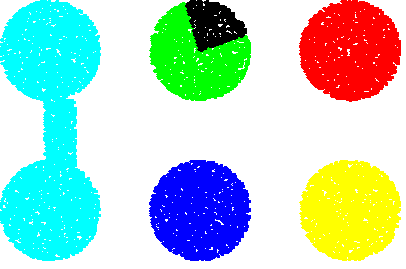} &
    \includegraphics[width=0.22\textwidth, height=0.18\textwidth]{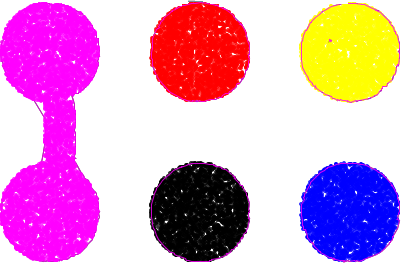} &
    \includegraphics[width=0.22\textwidth, height=0.18\textwidth]{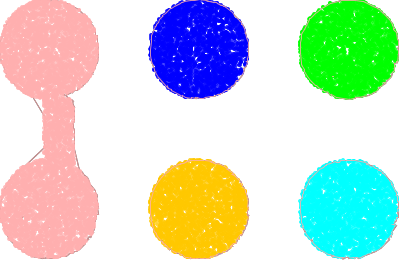} 
 \\ 
  \multicolumn{4}{c}{T2}
 \\ 
    \includegraphics[width=0.22\textwidth, height=0.18\textwidth]{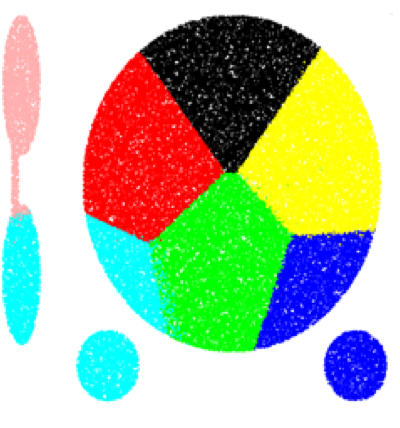} &
    \includegraphics[width=0.22\textwidth, height=0.18\textwidth]{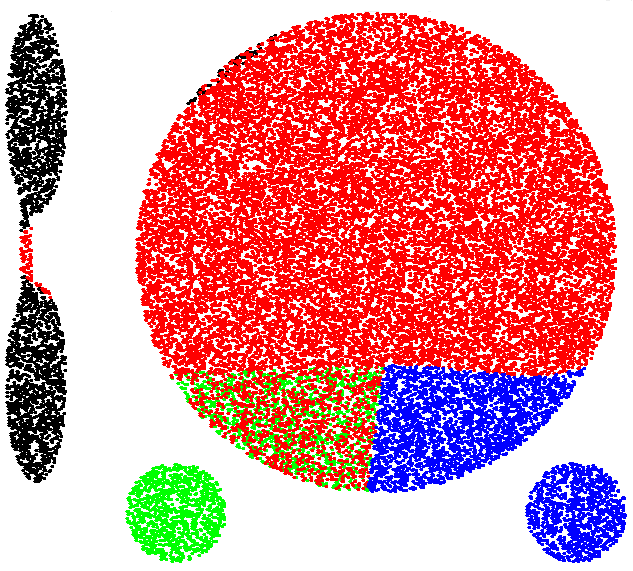} &
    \includegraphics[width=0.22\textwidth, height=0.18\textwidth]{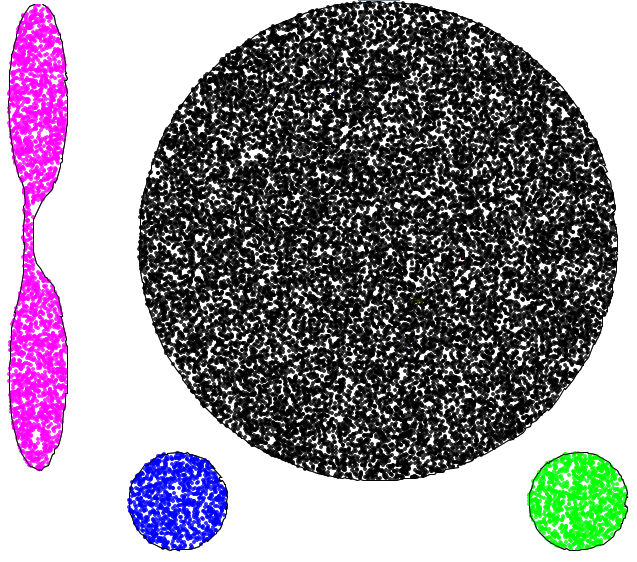} & 
    \includegraphics[width=0.22\textwidth, height=0.18\textwidth]{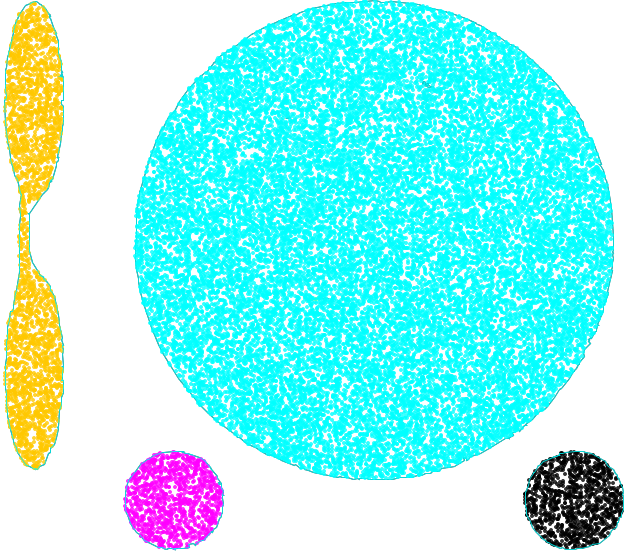} 
 \\ 
  \multicolumn{4}{c}{T3}
 \\   
\end{tabular}
\caption{Clusters generated for datasets ($T_1, T_2, T_3$).}
\label{SphSh}
\end{table*}

From the results shown in Figure \ref{SphSh}, as expected, since BIRCH cannot find all the
clusters correctly.  It splits the larger  cluster while merging the others.  In contrast,
CURE generates correctly the majority of the final clusters but it still fails to discover
all the clusters. Whereas both DDC-K-Means and DDC-DBSCAN algorithms successfully generate
all the clusters with the default parameter settings.

Figure  \ref{Nonesph} shows  the  clusters generated  for the  datasets  ($T_4, T_5,$  and
$T_6$). As expected, again BIRCH could not  find correct clusters; it tends to work better
with convex  shapes of the  clusters.  In  addition, BIRCH does  not deal with  noise. The
results of CURE are  worse and it is not able to extract  clusters with non-convex shapes.
We can also see that CURE does not deal with noise.  DDC-K-Means fails to find the correct
final results. In fact it returns the whole original dataset as one final cluster for each
dataset  ($T_4, T_5,$  and  $T_6$) (including  the  noise).  This  confirms  that the  DDC
technique is sensitive  to the type of  the algorithm chosen for the  first phase. Because
the  second phase  deals only  with the  merging of  the local  clusters whether  they are
correct or  not. This  issue is  corrected by  the DDC-DBSCAN,  as it  is well  suited for
non-convex shapes  of the  clusters and also  for eliminating the  noise and  outliers. In
fact, it generates good final clusters in datasets that have significant amount of noise.

As a final observation, these results prove  that the DDC framework is very efficient with
regard to  the accuracy  of its results.  The only  issue is to  choose a  good clustering
algorithm for the  first phase. This can  be done by exploring the  initial datasets along
with the question to be answered and choose a clustering algorithm accordingly.

Moreover, as for the DDC-K-Means, DDC-DBSCAN is dynamic (the correct number of clusters is
returned  automatically) and  efficient (the  approach  is distributed  and minimises  the
communications).

\subsection{Speed-up}

The goal here  is to study the execution  time of the four algorithms  and demonstrate the
impact of using a parallel and distributed  architecture to deal with the limited capacity
of a centralised system.

As mentioned in  Section \ref{DBCom}, the execution time for  the DDC-DBSCAN algorithm can
be generated in two cases. The first case  is to include the time required to generate the
distance matrix calculation.  The  second case is to suppose that  the distance matrix has
already been generated. The reason for this is that the distance matrix is calculated only
once.

\begin{table}[htb]
\caption{The execution times (ms) of BIRCH,  CURE, DDC-K-Means and DDC-DBSCAN with (w) and
  without (w/o) distance matrix computation.} 
\centering
\begin{tabular}{cr|r|r|r|r|r|}
\cline{3-7}
\multicolumn{2}{l}{} & \multicolumn{5}{|c|}{\textit{\textbf{Execution Time (ms)}}} \\ \cline{2-7} 
\multicolumn{1}{l|}{} & \multicolumn{1}{c|}{{\color[HTML]{000000} }} & \multicolumn{1}{c|}{} & \multicolumn{1}{c|}{}  & \multicolumn{1}{c|}{} & \multicolumn{2}{c|}{\textbf{DDC-DBSCAN}} \\ \cline{6-7} 
\multicolumn{1}{l|}{\multirow{-2}{*}{\textbf{}}} & \multicolumn{1}{c|}{\multirow{-2}{*}{{\color[HTML]{000000} \textbf{SIZE}}}} & \multicolumn{1}{c|}{\multirow{-2}{*}{\textbf{BIRCH}}} & \multicolumn{1}{c|}{\multirow{-2}{*}{\textbf{CURE}}} & \multicolumn{1}{c|}{\multirow{-2}{*}{\textbf{DDC-K-Means}}} & \multicolumn{1}{c|}{\textbf{W}} & \multicolumn{1}{c|}{\textbf{W/O}} \\ \hline
\multicolumn{1}{|c|}{\textbf{T1}} & 14000 & 328 & 145672 & 290 & 1049 & 632 \\ \hline
\multicolumn{1}{|c|}{\textbf{T2}} & 17080 & 312 & 405495 & 337 & 1283 & 814 \\ \hline
\multicolumn{1}{|c|}{\textbf{T3}} & 30350 & 347 & 1228063 & 501 & 1903 & 1220 \\ \hline
\multicolumn{1}{|c|}{\textbf{T4}} & 8000 & 249 & 72098 & 250 & 642 & 346 \\ \hline
\multicolumn{1}{|c|}{\textbf{T5}} & 10000 & 250 & 141864 & 270 & 836 & 470 \\ \hline
\multicolumn{1}{|c|}{\textbf{T6}} & 8000 & 218 & 92440 & 234 & 602 & 374 \\ \hline
\end{tabular}
\label{Table2}
\end{table}

Table \ref{Table2}  illustrates the execution  times of  the four techniques  on different
datasets. Note that the execution times do  not include the time for post-processing since
these are the same for the four algorithms.

As  mentioned in  Section  \ref{DBCom}, Table  \ref{Table2} confirmed  the  fact that  the
distance  matrix  calculation  in  DBSCAN is  very  significant.   Moreover,  DDC-DBSCAN's
execution time is  much lower than CURE’s  execution times across the  six datasets. Table
\ref{Table2} shows  also that  the DDC-K-Means  is very quick  which is  in line  with its
polynomial computational complexity. BIRCH is also  very fast, however, the quality of its
results  are not  good, it  failed in  finding  the correct  clusters across  all the  six
datasets.

The DDC-DBSCAN is a  bit slower that DDC-K-Means, but it returns  high quality results for
all the tested benchmarks, much better that DDC-K-Means, which has reasonably good results
for convex cluster shapes and very bad results for non-convex cluster shapes.  The overall
results confirm that  the DDC-DBSCAN clustering techniques compares favourably  to all the
tested  algorithms for  the  combined performance  measures (quality  of  the results  and
response time or time complexity).

\subsection{Scalability}

\begin{figure}[H]
\begin{center}
\includegraphics[width=\columnwidth]{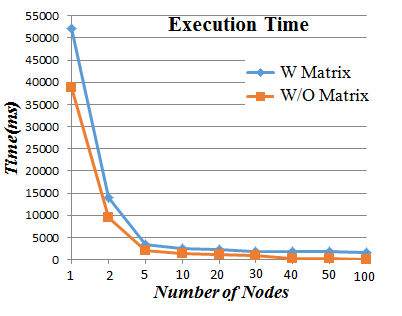}
\caption{Scalability Experiments.}
\label{SCAL}
\end{center}
\end{figure}

The goal  here is to determine  the effects of  the number of  nodes in the system  on the
execution times.  The  dataset contains $50,000$ data points. Figure  \ref{SCAL} shows the
execution time against the number of nodes ($x\_axis$ is in $log_2$) in the system. As one
can see,  DDC-DBSCAN took only  few seconds (including  the matrix computation's  time) to
cluster $50,000$ data points in a distributed  system that contains up to $100$ nodes, the
algorithm took  even less time  when we exclude the  matrix computation's time.  Thus, the
algorithm can comfortably handle high-dimensional data because of its low complexity.

\section{Conclusion}
\label{sec:Con}
In this paper, we proposed an efficient and flexible distributed clustering framework that
can work with existing  data mining algorithms.  The framework has  been tested on spatial
datsests using the K-Means and DBSCAN  algorithms.  The distributed clustering approach is
moreover dynamic, for spatial datasets, as it does  not need to give the number of correct
clusters  in  advance  (as  an  input  parameter). This  basically  solves  one  of  major
shortcomings  of K-Means  or DBSCAN.   The proposed  Distributed Dynamic  Clustering (DDC)
approach has  two main  phases: the  fully parallel phase  where each  node of  the system
calculates its own local clusters based on its  portion of the entire dataset. There is no
communications during  this phase, it takes  full advantage of task  parallelism paradigm.
The second  phase is also  distributed, but it  generates some communications  between the
nodes. However, the overhead due these communications has minimised by using a new concept
of cluster  representatives. Each cluster is  represented by its contour  and its density,
which count for about  $1\%$ of cluster size in general. This DDC  framework can easily be
implemented using MapReduce mechanism.

Note that,  as the first  phase is fully  parallel, each node  can use its  own clustering
algorithm  that suits  well its  local dataset.   However, the  quality of  the DDC  final
results depends heavily  on the local clustering used during  the first phase.  Therefore,
the issue  of exploring the original  dataset before choosing local  clustering algorithms
remains the key hurdle of all the data mining techniques.

The DDC approach was  tested using various benchmarks. The benchmarks  were chosen in such
a way to  reflect all the difficulties of clusters  extraction. These difficulties include
the shapes of the clusters (convex and non-convex), the data volume, and the computational
complexity. Experimental results  showed that the approach is very  efficient and can deal
with various situations (various shapes, densities, size, etc.).

As future  work, we  will study  in more details  the approach  scalability on  very large
datasets, and  we will explore  other models for  combining neighbouring clusters  such as
dynamic   tree-based   searching  and   merging   topology   on  large   scale   platforms
~\cite{Hudzia-05, savvas-04, N-A-09}. We will extend the framework to non-spatial datasets
by using Knowledge  Map ~\cite{N-Aa-07} for example.  We will also look at  the problem of
the data and communications reduction during phase two.

\section*{Acknowledgment}
\label{sec:Ack}
The  research work  is  conducted in  the  Insight  Centre for  Data  Analytics, which  is
supported by Science Foundation Ireland under Grant Number SFI/12/RC/2289.


%


\ifCLASSOPTIONcaptionsoff
  \newpage
\fi



%
\bibliography{DBSCANPaper}
\bibliographystyle{IEEEtran}

%


\begin{IEEEbiography}[{\includegraphics[width=1in,height=1.25in,clip,keepaspectratio]{picture}}]{John Doe}

\end{IEEEbiography}




\end{document}